\pdfoutput=1
\documentclass[sigconf, nonacm]{acmart}
\AtBeginDocument{%
  }
    
\setcopyright{none}
\copyrightyear{}
\acmYear{}
\acmDOI{}
\acmConference{}{}{}
\acmISBN{}

\usepackage{amsmath,amsthm}
\usepackage{graphicx}
\usepackage{booktabs}
\usepackage{hyperref}
\usepackage{xcolor}
\usepackage{caption}
\usepackage{subcaption}
\usepackage{tikz}
\usetikzlibrary{positioning}  
\usepackage{algorithm}
\usepackage{algpseudocode}   
\usepackage{enumitem}         
\usepackage{tabularx}
\usepackage{array}
\usepackage[table]{xcolor}
\usepackage{colortbl}
\usetikzlibrary{shapes,arrows,positioning,fit,calc}

\newtheorem{definition}{Definition}[section]

\begin{document}

\title{ReBound: Reuse-Aware Privacy For Interactive Decision Support}

\author{Nada Lahjouji}
\email{nlahjouj@uci.edu}
\affiliation{%
  \institution{University of California, Irvine}
  \city{Irvine}
  \country{USA}
}
\author{Shufan Zhang}
\email{shufan.zhang@uwaterloo.ca}
\affiliation{%
  \institution{University of Waterloo}
  \city{Waterloo}
  \country{Canada}
}

\author{Xi He}
\email{xi.he@uwaterloo.ca}
\affiliation{%
  \institution{University of Waterloo}
  \city{Waterloo}
  \country{Canada}
}

\author{Sharad Mehrotra}
\email{sharad@ics.uci.edu}
\affiliation{%
  \institution{University of California, Irvine}
  \city{Irvine}
  \country{USA}
}

\renewcommand{\shortauthors}{Interactive Privacy for Utility-Aware Decision Support}

\begin{abstract}
Differentially private decision support frameworks answer complex aggregate threshold queries with formal bounds on false negative and false positive rates, but treat each query independently with no memory of past results. In practice, analysts work interactively, issuing sequences of related queries that refine bounds, adjust thresholds, or derive new functions from previous ones. We propose \textsc{ReBound}, a framework that reuses cached results from previous queries to answer new queries at reduced or zero additional privacy cost while maintaining formal utility guarantees. \textsc{ReBound} introduces a reuse framework for multiple refinement types, a cache graph structure for efficient lookup of reusable results, and a negotiation mechanism for when requested bounds cannot be met within budget.
\end{abstract}

\maketitle

\section{Introduction}
Complex decision support (DS) queries with multiple aggregate conditions underpin many analytical applications, including clinical diagnosis~\cite{musen2021clinical,sutton2020overview}, building management~\cite{doukas2007intelligent}, and KPI evaluation~\cite{dombrowski2013kpi}. They commonly arise in online analytical processing (OLAP)~\cite{chaudhuri1997olap}, where analysts compute group-level aggregates and compare them to thresholds to drive decisions. However, such data sources often contain sensitive individual information, meaning aggregate releases can cause serious privacy leaks~\cite{dinur2003revealing,dwork2008disclosure}. Differential privacy (DP)~\cite{dwork2006calibrating,dwork2014algorithmic} addresses this by limiting any single record's influence on the output, but introduces noise that can harm decision quality. In DS settings, mechanisms must therefore provide formal \emph{utility} guarantees: noisy answers can induce false positives (FP), where predicates are incorrectly accepted, and false negatives (FN), where true predicates are missed. This motivates a ``utility-first'' line of work~\cite{ligett2017accuracy,ge2019apex,lobo2020dpella,ghayyur2022mide,lahjouji2024probe} that specifies target utility and maximizes privacy subject to that requirement, rather than fixing a privacy budget and optimizing utility.

Prior work has made significant progress on accuracy-first differential privacy for decision support.
APEx~\cite{ge2019apex} introduced the paradigm but leaves errors unbounded near thresholds;
MIDE~\cite{ghayyur2022mide} provides one-sided FNR guarantees for simple queries; and 
ProBE~\cite{lahjouji2024probe} achieves dual $(\alpha, \beta)$ guarantees for complex queries with minimal privacy loss. However, all three treat each query \emph{independently} and do not exploit the natural overlap in interactive analyst sessions.

In practice analysts rarely issue queries in isolation. The nature of decision support is inherently \emph{interactive}: an analyst issues an initial query, inspects the results, and then refines their analysis by tightening accuracy bounds, adjusting thresholds, changing logical operators, or combining previously queried metrics. Under the current paradigm, each such refinement incurs the full privacy cost as if it were an entirely new query, causing the cumulative privacy budget to grow linearly with the number of interactions even when successive queries are closely related. Consider, for instance, the following sequence over a medical dataset:
\begin{table}[h]
\centering
\caption{Example query sequence ($Q_1, Q_2, Q_3$) treated as independent with privacy budget $\epsilon_1, \epsilon_2, \epsilon_3$ respectively.}
\label{tab:dp_running_example}
\small
\setlength{\tabcolsep}{2.5pt}
\renewcommand{\arraystretch}{1.0}

\begin{tabularx}{\columnwidth}{@{}>{\raggedright\arraybackslash}X
                                p{0.1\columnwidth}
                                p{0.2\columnwidth}@{}}
\toprule
\textbf{Query} & $\boldsymbol{\beta}$ & \textbf{Cumulative Privacy Cost} \\
\midrule
\textbf{Q1:} COUNT(Flu)$>100$ AND AVG(Temp)$>101$
& 0.1 & $\epsilon_1$ \\
\midrule
\textbf{Q2:} $\equiv$ Q1 (same query, tighter $\beta$)
& 0.05 & $\epsilon_1+\epsilon_2$ \\
\midrule
\textbf{Q3:} COUNT(Flu)$>120$ AND AVG(Temp)$>103$ (same predicates, different thresholds)
& 0.05 & $\epsilon_1+\epsilon_2+\epsilon_3$ \\
\bottomrule
\end{tabularx}
\end{table}


\noindent Yet the noisy aggregate values computed for Q1 already contain information that could be leveraged to answer Q2 (the same query with a tighter bound) and Q3 (which only differs in the thresholds) at a significantly reduced cost. This observation motivates the central question of this work: \emph{how can we systematically reuse cached results from previous differentially private queries to answer new, related queries while maintaining formal FNR and FPR guarantees?}
Answering this question introduces several challenges.

First, different types of query refinements admit fundamentally different reuse strategies. Tightening the accuracy bound $\beta$ on the same query, shifting the threshold $c$, changing the logical structure (AND $\leftrightarrow$ OR), and deriving new aggregate functions from cached ones each require distinct derivation with different privacy cost implications. A systematic framework to identify \emph{when} reuse is possible and \emph{how} to exploit it for each type of refinement is thus needed. 
Second, the cache must store not only final query answers but also intermediate results at the right granularity to maximize reuse opportunities. For instance, caching the results of composed sub-expressions (e.g., $Q_1 \cap Q_2$) in addition to individual atomic sub-query results enables reuse even when the overall query structure changes. The design of this multi-layered cache structure, including what to store and how to efficiently identify maximal overlap with incoming queries, is a non-trivial design problem.
Third, when a requested query cannot fully be answered within the remaining privacy budget, the system should be able to propose alternative queries (e.g., with relaxed bounds or adjusted thresholds) that can be answered at lower cost, presenting the analyst with meaningful trade-offs rather than simply denying the query.

In this paper, we propose \textsc{ReBound}, a framework for differentially private interactive decision support that addresses the three challenges above. At its core, \textsc{ReBound} proposes a reuse framework that supports multiple types of query changes, including threshold changes, bound tightening, logical structure modifications, and linear combinations of cached aggregate functions, with minimal additional privacy loss while maintaining formal FNR and FPR guarantees. To do so, \textsc{ReBound} uses a cache graph structure that stores not only atomic sub-query results but also composed sub-expressions, enabling efficient lookup of maximal reusable subgraphs when a new query arrives and fast insertion of new results upon execution. Finally, \textsc{ReBound} supports a privacy negotiation mechanism wherein queries requesting accuracy bounds that cannot be met within the remaining privacy budget result in a counteroffer with relaxed bounds or adjusted parameters that can be satisfied at lower cost, allowing the analyst to accept an alternative rather than receiving an outright denial.


\section{Background}

Let $D \in \mathcal{D}$ denote a sensitive dataset. Two datasets $D,D'\in\mathcal{D}$ are \emph{neighbors}, written $D \sim D'$, if they differ in one tuple. We consider the following \emph{ex-post} differential privacy notion.

\begin{definition}[Ex-Post Differential Privacy]
Let $\mathcal{M}:\mathcal{D}\to \mathcal{O}$ be a randomized mechanism and
$\mathcal{E}:\mathcal{O}\to [0,\infty]$ be an outcome-dependent privacy bound.
For an output $o\in\mathcal{O}$, define the ex-post privacy loss as
\[
\mathcal{E}(o)\triangleq \max_{D\sim D'} \ln \frac{\Pr[\mathcal{M}(D)=o]}{\Pr[\mathcal{M}(D')=o]}.
\]
We say $\mathcal{M}$ satisfies \emph{$\mathcal{E}(o)$-ex-post DP} if for all $o\in\mathcal{O}$, $\varepsilon(o) \le \mathcal{E}(o)$.
\end{definition}

A complex decision-support (DS) query $Q_{\Lambda,F}$ is an expression over atomic queries $Q_{a_1},\ldots,Q_{a_n}$ that share the same predicate set $\Lambda$ but may have different filters $F=\{f_1,\ldots,f_n\}$, aggregates $G=\{g_1,\ldots,g_n\}$, and thresholds $C=\{c_1,\ldots,c_n\}$, combined using $\cap$ (AND) and $\cup$ (OR). Equivalently:
\[
Q_{\Lambda,F} ::= Q_a \mid Q_{\Lambda,F_1}\cap Q_{\Lambda,F_2} \mid Q_{\Lambda,F_1}\cup Q_{\Lambda,F_2}.
\]
Each atomic aggregate threshold query is defined as follows.
\begin{definition}[Atomic Aggregate-Threshold Query~\cite{ge2019apex,lahjouji2024probe}]
Given $(\Lambda,f,g,C)$, the atomic decision-support query returns the set of predicates
whose group-wise aggregate exceeds the threshold:
\[
Q_a(D;\Lambda,f,g,C) \triangleq \big\{\lambda\in\Lambda : g(D_{f,\lambda}) > c_\lambda \big\}.
\]
\end{definition}

The accuracy constraints for DS queries are specified as predicate-wise false negative rates (FNRs) and false positive rates (FPRs).

\begin{definition}[FNR/FPR Bounds]
Let $\widehat{A}\subseteq \Lambda$ denote the (random) output returned by a DP mechanism.
For each $\lambda\in\Lambda$, define
\[
\mathrm{FN}_\lambda \triangleq \mathbf{1}[\lambda\in A \wedge \lambda\notin \widehat{A}],\qquad
\mathrm{FP}_\lambda \triangleq \mathbf{1}[\lambda\notin A \wedge \lambda\in \widehat{A}].
\]
A DS request specifies utility parameters $(\alpha,\beta)\in[0,1]^2$ requiring
\[
\forall \lambda\in\Lambda:\quad
\Pr[\mathrm{FN}_\lambda = 1] \le \alpha
\quad \text{and}\quad
\Pr[\mathrm{FP}_\lambda = 1] \le \beta.
\]
\end{definition}

\section{Problem Setup}
We consider an \textbf{interactive} decision-support workflow where an analyst \emph{adaptively} issues DS queries, possibly depending on prior answers.
At each round, the system minimizes incremental privacy cost while meeting $(\alpha,\beta)$ under a budget cap $\varepsilon_{\max}$; if infeasible, it proposes a minimal relaxation of $(\alpha,\beta)$.
Under this interactive setting, several interesting research questions arise.

\noindent
\textbf{RQ1: Designing DP Caches for DS Queries.}
Existing work~\cite{mazmudar2022cache,ZhangH23dprovdb,KostopoulouTCGL23turbo} on DP caching all focus on linear/counting workloads, and store historical noisy answers/parameters to reuse them under post-processing.
Their target accuracy notions are primarily numerical error bounds. However, DS query workloads~\cite{ge2019apex,ghayyur2022mide,lahjouji2024probe} are threshold-classification queries where the utility is often defined as predicate-wise FNRs and/or FPRs.
The outputs of DS queries are sets rather than numerical results, making it insufficient to just store the noisy answers and recheck the bounds when analysts change the query thresholds or demand different FNR/FPR guarantees.

Our goal is to design \emph{efficient} DP caches for DS queries that capture not only answers to existing queries, but an indexed region of future DS queries for which the answers can be reused and composed.
We would like to develop a novel hierarchical structure that organizes domain information, query metadata, decomposed predicates, and historical query answers, alongside algorithms to efficiently look up and update the cache when answering DS queries.

\noindent
\textbf{RQ2: Cache-Aware Privacy Optimization under Budget Constraints.}
Given the nature of uncertainty in the online/interactive query systems, an arriving complex DS query may partially overlap with historical queries yet differ in several ways, e.g., tighter FNR/FPR requirements, modified logical structure, shifted predicate thresholds, or even newly introduced predicates not present in the cache.
When the cache cannot fully answer the request, how can the system acquire only the missing information and combine it with cached DP results to satisfy $(\alpha,\beta)$ with minimal incremental privacy cost, or else compute a minimal feasible utility bound relaxation under $\varepsilon_{\max}$?

Prior work~\cite{PrivacyRelax,DBLP:conf/nips/WhitehouseRWR22,ZhangH23dprovdb} on refining accuracy guarantees by calibrating correlated noise and adaptive reuse is largely tailored to numerical error metrics and fixed query semantics, and therefore does not directly extend to DS queries.
APEx~\cite{ge2019apex} optimizes the per-query budget for DS queries, but it treats each arriving query as fresh (no cache); moreover, the uniform privacy apportionment assumptions (across predicates) of its mechanism can be misaligned with heterogeneous cached uncertainties, leading to suboptimal ``top-ups'' when only parts of a query can be answered from cached results.
In this paper, we design cache-aware DP mechanisms that certify reusable sub-structures under new utility constraints and minimize incremental privacy cost via structure-aware allocation. When infeasible under $\varepsilon_{\max}$, the mechanisms should return a minimal $(\alpha,\beta)$ relaxation as a counteroffer.

\begin{figure}[h]
    \centering
    \includegraphics[width=0.89\columnwidth]{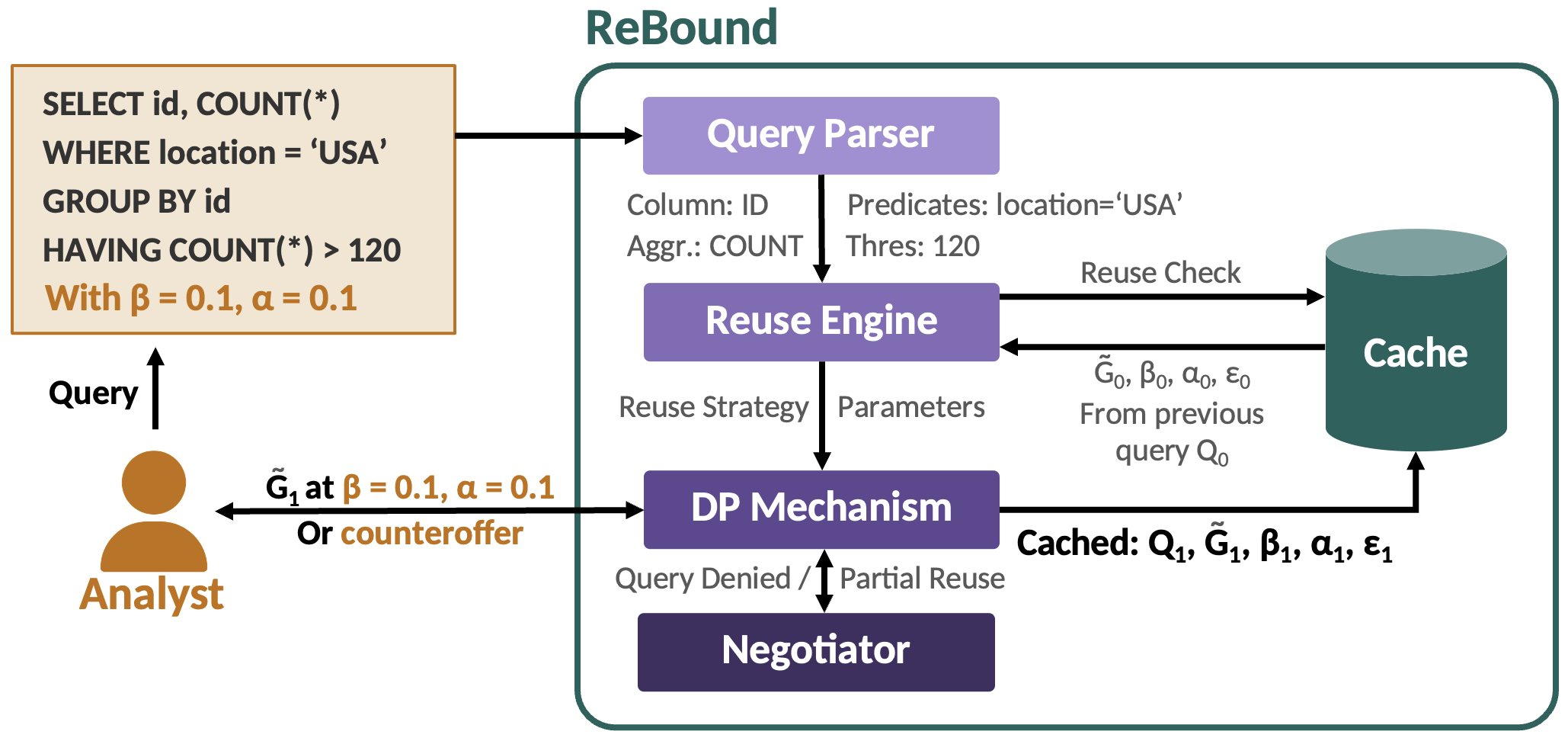}
    \caption{System Overview of \textsc{ReBound}.}
    \label{fig:overview}
\end{figure}


 \vspace{-0.4cm}

\section{Framework Overview}
\label{sec:overview}

Interactive DP workloads exhibit natural query overlap: analysts refine predicates, adjust thresholds, and explore related aggregates over time.
\textsc{ReBound} maintains a structured cache of prior DP releases and derives new answers from cached results when possible, negotiating with analysts when budget constraints conflict with accuracy requirements (Figure~\ref{fig:overview}).
Our design follows three principles: \textbf{post-processing reuse}, where computations over prior DP outputs incur no additional privacy cost; \textbf{utility propagation}, where derived results have computable $(\alpha,\beta)$ bounds that are verified before returning reused answers; and \textbf{budget-aware negotiation}, where the system and analyst negotiate acceptable tradeoffs when reuse is insufficient under a constrained budget.
We build on ProBE~\cite{lahjouji2024probe} as a base mechanism, which guarantees $(\alpha, \beta)$ bounds at a minimal privacy loss, but treats each query independently. It first decomposes complex queries into atomic sub-queries and optimally apportions $\beta$ across them~\cite{ghayyur2022mide} to guarantee the overall $\beta$-bound, then enforces $\alpha$ in a data-dependent step that estimates false positives and tightens the error region around the threshold when needed.

\noindent The remainder of this section describes reuse strategies, cache structure, and query processing; formal proofs are deferred to the full paper.

\vspace{-0.2cm}
\subsection{Reuse Strategies}
\label{sec:reuse}
\textsc{ReBound} treats reuse as a candidate derivation from cached DP releases and returns an answer at $\Delta\varepsilon=0$ only when the requested utility bounds are met. For each \texttt{HAVING} atom $(\phi,\mathcal{F},\tau)\in\mathcal{A}$, it traverses cached predicate nodes $\{P_i\}$ and scalar nodes $\{A_i\}$ to derive a candidate by post-processing, propagates a $\beta$ bound from the contributing releases, and then checks whether the induced false-positive guarantee satisfies $\alpha$. If the $\alpha$ check fails, \textsc{ReBound} runs ProBE~\cite{lahjouji2024probe}'s $\alpha$-guarantee step, spending incremental cost $\varepsilon' \! >0$ to tighten the uncertainty region. We defer proofs of the propagation procedure and associated parameters to the full paper.

\noindent\textbf{Exact Match.}
If the cache contains a release for the same atom $(\phi,\mathcal{F},\tau)$, \textsc{ReBound} reuses the corresponding scalar node $S_i$ and its noisy value $\widetilde{G}$ by post-processing.
The candidate inherits $\beta$ from the cached release, after which \textsc{ReBound} checks whether the induced $\alpha$ meets the requested bound.
If so, reuse is returned with $\Delta\varepsilon=0$; otherwise, \textsc{ReBound} runs \cite{lahjouji2024probe}'s $\alpha$-step starting from the cached state, incurring incremental cost $\varepsilon'$ to satisfy $\alpha$.

\noindent\textbf{Threshold Change.}
For the same $(\phi,\mathcal{F})$ but a different threshold $\tau'$, \textsc{ReBound} reuses the cached noisy scalar $\widetilde{G}$ and recomputes the decision $\mathbb{I}[\widetilde{G}>\tau']$ by post-processing.
The propagated $\beta$ bound transfers from the cached release (since it depends on the uncertainty region rather than the threshold), but the false-positive bound must be re-checked for $\tau'$.
If the $\alpha$ requirement is not met, \textsc{ReBound} again  spends incremental budget $\varepsilon'$ to satisfy $\alpha$ (or reuse is rejected).

\noindent\textbf{Predicate Decomposition.}
If a requested predicate $\phi$ can be expressed using cached predicate nodes $\{P_i\}$ via set operations (e.g., union, intersection, set difference), \textsc{ReBound} derives the corresponding aggregate by combining the relevant cached scalars.
For additive functions (\texttt{COUNT}, \texttt{SUM}), these set operations induce linear combinations of noisy scalars (e.g., disjoint unions sum).
\textsc{ReBound} propagates $\beta$ conservatively across the contributing releases and then checks whether the derived decision satisfies the requested $\alpha$, after which it 
may again incur an incremental cost $\varepsilon'$ to ensure the final FPR is bounded by $\alpha$.

\noindent\textbf{Linear Combination.}
If a requested aggregate can be expressed as a deterministic function of cached scalar nodes $\{S_i\}$ (e.g., linear forms for additive aggregates like \texttt{COUNT(x)}+\texttt{COUNT(y)}, or rewriting ratios like $\texttt{AVG}=\texttt{SUM}/\texttt{COUNT}$), \textsc{ReBound} derives the candidate value by post-processing and propagates $\beta$ from its inputs by ensuring the combined aggregate uses the appropriate derived parameters (e.g. sensitivity and error region).
The $\alpha$ bound is then enforced on the combined aggregate result as a single threshold test, not on individual sub-queries. 






\subsection{Cache Structure}
\label{sec:cache}

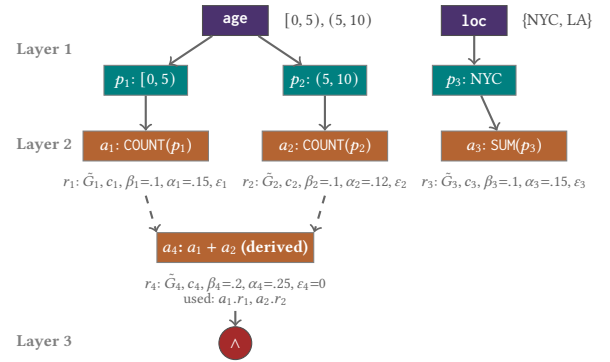
\begin{figure}[h]
    \centering
    \small
    \begin{tikzpicture}[
        scale=0.79, transform shape,
        node distance=0.4cm and 0.6cm,
        every node/.style={font=\small},
        colnode/.style={rectangle, draw=black!70, fill={rgb,255:red,75;green,50;blue,110}, text=white, minimum width=1.1cm, minimum height=0.5cm, font=\small\bfseries},
        prednode/.style={rectangle, draw=black!50, fill={rgb,255:red,0;green,128;blue,128}, text=white, minimum width=1.4cm, minimum height=0.5cm},
        scalarnode/.style={rectangle, draw=black!50, fill={rgb,255:red,180;green,100;blue,50}, text=white, minimum width=2.1cm, minimum height=0.5cm},
        derivednode/.style={rectangle, draw=black!50, fill={rgb,255:red,180;green,100;blue,50}, text=white, minimum width=1.9cm, minimum height=0.5cm, font=\small\bfseries},
        boolnode/.style={circle, draw=black!70, fill={rgb,255:red,165;green,42;blue,42}, text=white, minimum size=0.55cm, font=\small\bfseries},
        layerlabel/.style={font=\small\bfseries, text=gray},
        intervals/.style={font=\small, text=black!70},
        release/.style={font=\footnotesize, text=black!60, align=center}
    ]

    \node[layerlabel] at (-4.2, 0.1) {Layer 1};
    \node[layerlabel] at (-4.2, -1.5) {Layer 2};
    \node[layerlabel] at (-4.2, -4.8) {Layer 3};

    \node[colnode] (age) at (-1.0, 0.6) {\texttt{age}};
    \node[colnode] (loc) at (3.0, 0.6) {\texttt{loc}};

    \node[intervals, right=0.15cm of age] {\small$[0,5),(5,10)$};
    \node[intervals, right=0.15cm of loc] {\{NYC, LA\}};

    \node[prednode] (p1) at (-2.5, -0.4) {$p_1$: $[0,5)$};
    \node[prednode] (p2) at (0.5, -0.4) {$p_2$: $(5,10)$};
    \node[prednode] (p3) at (3.0, -0.4) {$p_3$: NYC};
    
    \draw[->, thick, black!60] (age) -- (p1);
    \draw[->, thick, black!60] (age) -- (p2);
    \draw[->, thick, black!60] (loc) -- (p3);

    \node[scalarnode] (a1) at (-2.5, -1.5) {$a_1$: \texttt{COUNT}($p_1$)};
    \node[scalarnode] (a2) at (0.5, -1.5) {$a_2$: \texttt{COUNT}($p_2$)};
    \node[scalarnode] (a3) at (3.5, -1.5) {$a_3$: \texttt{SUM}($p_3$)};
    
    \draw[->, thick, black!60] (p1) -- (a1);
    \draw[->, thick, black!60] (p2) -- (a2);
    \draw[->, thick, black!60] (p3) -- (a3);
    
    \node[release, below=0.08cm of a1] (r1) {$r_1$: $\tilde{G}_1, c_1, \beta_1{=}.1, \alpha_1{=}.15, \varepsilon_1$};
    \node[release, below=0.08cm of a2] (r2) {$r_2$: $\tilde{G}_2, c_2, \beta_2{=}.1, \alpha_2{=}.12, \varepsilon_2$};
    \node[release, below=0.08cm of a3] (r3) {$r_3$: $\tilde{G}_3, c_3, \beta_3{=}.1, \alpha_3{=}.15, \varepsilon_3$};

    \node[derivednode] (a4) at (-1.0, -3.2) {$a_4$: $a_1 + a_2$ (derived)};

    \draw[->, dashed, thick, black!60] (r1.south) -- (a4.north west);
    \draw[->, dashed, thick, black!60] (r2.south) -- (a4.north east);

    \node[release, below=0.08cm of a4] (r4) {$r_4$: $\tilde{G}_4, c_4, \beta_4{=}.2, \alpha_4{=}.25, \varepsilon_4{=}0$\\[-0.3ex] used: $a_1.r_1, a_2.r_2$};
    
    \node[boolnode] (b1) at (-1.0, -4.8) {$\wedge$};
    \draw[->, thick, black!60] (r4.south) -- (b1.north);
    
    \end{tikzpicture}
    \caption{Cache structure with example nodes/releases.}
    \label{fig:cache}
\end{figure}

To efficiently identify reuse opportunities, the cache organizes prior releases as a three-layer DAG (Figure~\ref{fig:cache}) reflecting query anatomy.

\noindent\textbf{Layer 1: Predicate Index.}
Column nodes represent \texttt{GROUP BY} attributes; predicate nodes store intervals (numeric) or value sets (categorical). A boundary registry enables predicate decomposition.

\noindent\textbf{Layer 2: Scalar Aggregates.}
Each scalar node (e.g., $a_1$, $a_2$, $a_3$) corresponds to a (predicate, function) pair and maintains releases under different $(c, \beta, \alpha, \varepsilon)$ parameters. Derived nodes (e.g., $a_4 = a_1 + a_2$) store derivation expressions and source references.

\noindent\textbf{Layer 3: Boolean Composition.}
Represents compound \texttt{HAVING} clauses (\texttt{AND}/\texttt{OR}), enabling reuse under composition changes.

\begin{definition}[Cache Structure]
\label{def:cache}
The cache $\mathcal{C} = (V, E, R)$ is a three-layer DAG where $V = V_C \cup V_P \cup V_S \cup V_B$ are column, predicate, scalar, and boolean nodes; $E$ encodes cross-layer derivability; and $R: V_S \to 2^{\mathcal{R}}$ maps scalar nodes to privacy releases $r = (\tilde{G}, c, \beta, \alpha, \varepsilon)$ representing previously released noisy values.
\end{definition}

\vspace{-0.2cm}
\subsection{Query Processing}
\label{sec:lifecycle}

\begin{algorithm}[h]
\caption{Query Processing in \textsc{ReBound}}
\label{alg:query}
\small
\begin{algorithmic}[1]
\Require Query $Q$, utility $\alpha,\beta$, budget limit $\varepsilon_{\max}$
\State $(\Phi,\mathcal{A},B)\gets \textsc{Decompose}(Q)$
\Comment{$\Phi=\{\phi_i\}$ predicates; $\mathcal{A}=\{(\phi_i,\mathcal{F},\tau)\}$ aggregates; $B$ is boolean nodes}
\State $S^* \gets \textsc{ChooseStrategy}(\Phi,\mathcal{A},B,\mathcal{C},\beta,\alpha,\varepsilon_{\max})$
\If{$\textsc{ReuseOnly}(S^*)$}
    \State \Return $\textsc{DeriveFromCache}(\Phi,\mathcal{A},B,S^*)$ \Comment{$\varepsilon = 0$}
\EndIf
\State $(ans,\varepsilon_{\text{actual}}) \gets \textsc{PROBE}(\Phi,\mathcal{A},B,S^*,\beta,\alpha)$
\If{$\varepsilon_{\text{actual}} \le \varepsilon_{\max}$}
    \State $\textsc{CacheInsert}(ans)$; \Return $ans$
\Else
    \State \Return $\textsc{Negotiate}(Q,\beta,\alpha,\varepsilon_{\text{actual}},\varepsilon_{\max})$
\EndIf
\end{algorithmic}
\end{algorithm}


Algorithm~\ref{alg:query} outlines query processing. Given an incoming query $Q$ with utility requirements $U=(\alpha,\beta)$ and budget limit $\varepsilon_{\max}$, \textsc{ReBound} first parses $Q$ into a canonical form capturing the \texttt{GROUP BY} key, a predicate set $\Phi=\{\phi_i\}$, and a set of aggregate atoms $\mathcal{A}=\{(\phi_i,\mathcal{F},\tau)\}$ used in the \texttt{HAVING} clause (lines~1--2), where $\phi_i$ is a (group-level) selection predicate, $\mathcal{F}$ denotes an aggregate function (e.g., \texttt{COUNT}), and $\tau$ is the corresponding threshold. It then selects a reuse strategy $S^*$ by traversing the cache DAG layer-by-layer (line~3). At the predicate layer, for each $\phi\in\Phi$, the strategy checks whether its record set can be expressed using cached predicate nodes $\{P_i\}$ via set operations (e.g., union or set difference), and records a concrete decomposition when possible. At the scalar-aggregate layer, for each requested atom in $\mathcal{A}$, the strategy checks whether the required noisy scalar can be derived from cached aggregate nodes $\{A_i\}$ by post-processing (e.g., $\texttt{AVG}=\texttt{SUM}/\texttt{COUNT}$). If the traversal yields a complete derivation, \textsc{ReBound} returns the answer via post-processing at $\varepsilon=0$ (lines~4--5); otherwise it executes \textsc{PROBE}~\cite{lahjouji2024probe} for only the unresolved components according to $S^*$, returning the final answer and realized cost $\varepsilon_{\text{actual}}$ (line~6).

\noindent\textbf{Negotiation.}
If $\varepsilon_{\text{actual}} > \varepsilon_{\max}$, negotiation is triggered: the system offers relaxed accuracy bounds (either through minimal relaxation of $\beta$ or through an effective upper bound $\alpha$) yielding lower $\varepsilon$, which the analyst can accept or deny. Negotiation may also handle \emph{partial reuse}: queries not fully matching any reuse case but partially derivable from the cache. We leave the formalization of the negotiation mechanism to the full paper.

\noindent\textbf{Guarantee.}
\textsc{ReBound} guarantees that every query answer satisfies $\varepsilon_{\max}$-DP with a $\beta$-bound on FNR and $\alpha$-bound on FPR when possible. When the requested bounds cannot be achieved under $\varepsilon_{\max}$-DP, the negotiation framework suggests next best bounds $\beta$ and $\alpha$.
\vspace{-0.1cm}

\section{Preliminary Results}
\begin{table}[h]
\centering
\caption{Analyst session queries used over the NYC Taxi dataset. Each pattern contains 10 queries total.}
\label{tab:workload}
\scriptsize
\setlength{\tabcolsep}{3pt}
\renewcommand{\arraystretch}{1.05}

\begin{tabularx}{\columnwidth}{@{}p{1.65cm} >{\raggedright\arraybackslash}X p{1.25cm}@{}}
\toprule
\textbf{Phase} & \textbf{Query Example} & \textbf{Change} \\
\midrule

\rowcolor{gray!15}
\multicolumn{3}{@{}l}{\textbf{\textit{Pattern A: Drill-and-Tighten}}} \\[1pt]
Seed (Q1) &
\texttt{CNT(*)} $\wedge$ \texttt{AVG(tip)}, $\beta{=}0.10$, $\alpha{=}0.20$ & -- \\
Bound (Q2--Q7) &
Same query, progressively tighten $\beta, \alpha$ & $\beta/\alpha$ \\
Thresh.\ (Q8--Q10) &
Same, threshold $\times 1.10$, $\times 1.15$, $\times 1.20$ & Threshold \\[2pt]

\rowcolor{gray!15}
\multicolumn{3}{@{}l}{\textbf{\textit{Pattern B: Exploratory Branching}}} \\[1pt]
Seed (Q1) &
\texttt{AVG(tip)}, $\beta{=}0.05$, $\alpha{=}0.15$ & -- \\
Pred.\ (Q2, Q4) &
\texttt{AVG(tolls)}; \texttt{AVG(congestion)} & Predicate \\
Sum (Q3, Q5, Q9) &
\texttt{AVG(tip) + AVG(tolls)}; \texttt{AVG(tip) + AVG(congestion)} & Linear comb. \\
Conj.\ (Q6--Q8, Q10) &
\texttt{CNT(fare>10)} $\wedge$ \texttt{AVG(tip)}; \texttt{CNT(fare<5)} $\wedge$ \texttt{AVG(tip)} & Agg./Pred. \\

\bottomrule
\end{tabularx}
\end{table}

To evaluate \textsc{ReBound}, we construct analyst sessions inspired by OLAP session templates~\cite{rizzi2014cubeload}, where each query incrementally refines the previous one (e.g., drill-down, slice, or measure change). We instantiate two 10-query session patterns over the NYC Taxi dataset~\cite{nytaxi} ($\sim$3 million trips from March 2020), grouping by hashed pickup location and using \texttt{COUNT} and \texttt{AVG} aggregates under varying predicates, thresholds, and utility targets. Table~\ref{tab:workload} summarizes the patterns: \textbf{Pattern~A} (Drill-and-Tighten) progressively tightens $(\alpha,\beta)$ and then adjusts thresholds, while \textbf{Pattern~B} (Exploratory Branching) pivots across predicates, aggregates, and linear combinations. We compare cumulative privacy cost under \textsc{ReBound} against a cache-less baseline using sequential composition, averaging results over 10 runs to account for Laplace noise.

\begin{figure}
\vspace{-2mm}
    \centering
    \includegraphics[width=0.65\linewidth]{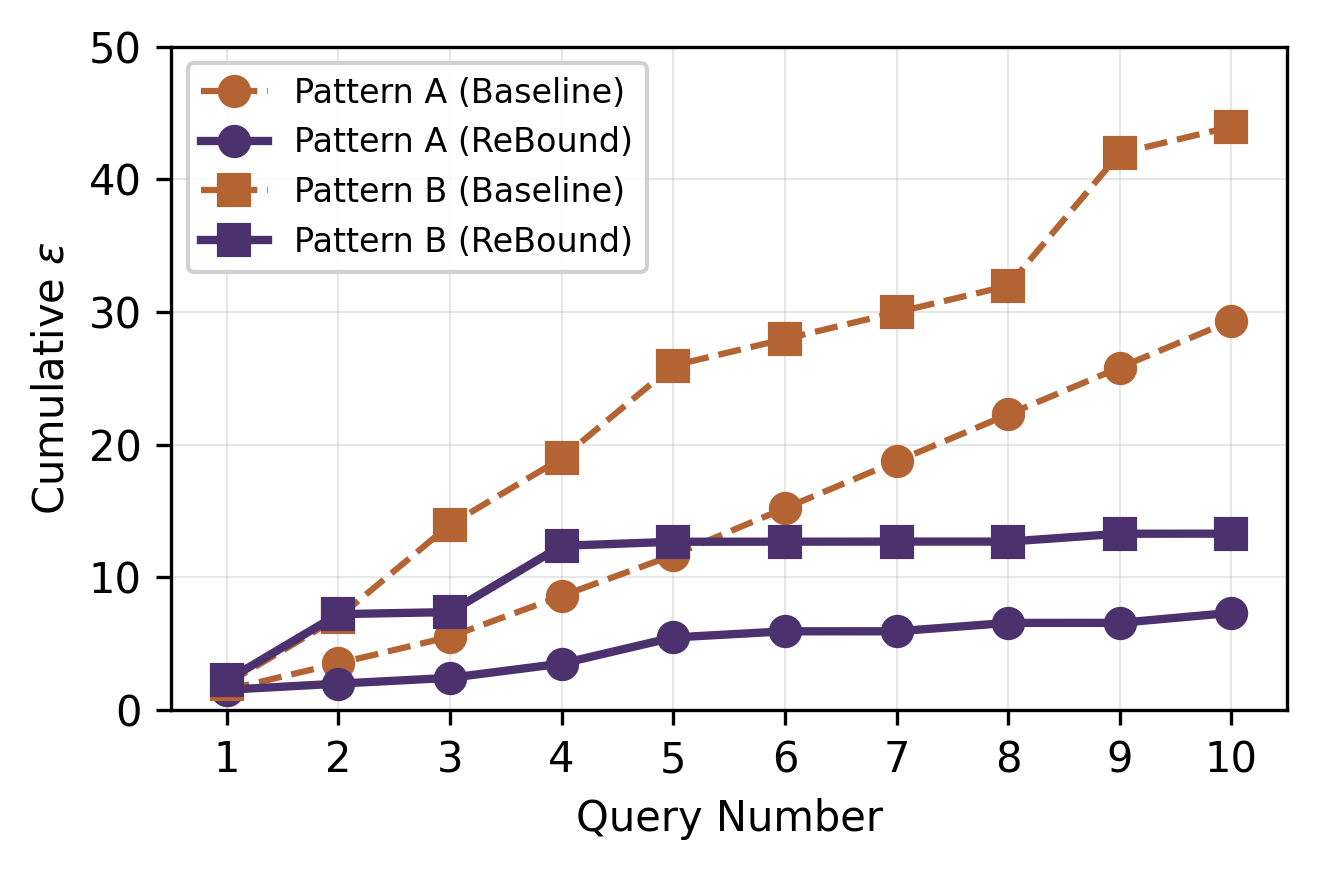}
    \caption{Cumulative privacy loss over query sessions. Dashed orange lines show baseline (no reuse); solid purple lines show \textsc{ReBound}.}
    \label{fig:cumulative}
\end{figure}

\noindent\textbf{Privacy Savings.}
Figure~\ref{fig:cumulative} shows cumulative privacy loss over each query session.
\textsc{ReBound} achieves substantial savings by reusing cached releases: 75\% reduction for Pattern A and 70\% for Pattern B.
The gap widens as sessions progress, proving that later queries benefit from richer cache state built by earlier queries. More importantly, these savings translate into longer interactive analysis sessions under fixed privacy budgets: under fixed session budgets, the baseline answers only 4 and 3 queries, respectively, while \textsc{ReBound} completes all 10 in both cases.

\begin{table}[h]
\centering
\small
\caption{Queries completed under fixed privacy budgets.}
\label{tab:queries}
\setlength{\tabcolsep}{4pt}
\renewcommand{\arraystretch}{1.05}

\begin{tabular}{lcc}
\toprule
\textbf{Pattern} & \textbf{Baseline} & \textbf{\textsc{ReBound}} \\
\midrule
A ($\varepsilon_{\max}=10$) & 4 / 10 & \textbf{10 / 10} \\
B ($\varepsilon_{\max}=15$) & 3 / 10 & \textbf{10 / 10} \\
\bottomrule
\end{tabular}
\end{table}

\noindent\textbf{Negotiation.} We evaluate bound-based negotiation (relaxing $\alpha$ or $\beta$); partial reuse negotiation is left for future work.
Without negotiation, Pattern A resulted in 12 query denials across 10 runs: Q7 was denied in 9/10 runs and Q9 in 3/10 runs, both requesting tight bounds ($\beta = 0.005$, $\alpha = 0.05$) with all denials happening due to the $\alpha$ bound.
With negotiation enabled, all queries succeeded.
The system offered relaxed $\alpha$ bounds of 0.063 (Q7) and 0.054 (Q9), suggesting minor relaxations from the requested 0.05, while $\beta$ remained effectively unconstrained under a $\varepsilon_{\max} = 9$ budget.

\vspace{-0.1cm}
\section{Concluding Remarks}
We presented \textsc{ReBound}, a cache-aware framework for interactive differentially private decision support.
\textsc{ReBound} maintains a structured cache of prior DP releases and applies formally defined reuse conditions to answer follow-on queries via post-processing when possible, reducing incremental privacy loss and enabling more queries under a fixed budget; when reuse is insufficient, it negotiates accuracy-budget tradeoffs to maximize the answered workload.
In future work, \textsc{ReBound} will be implemented as a fully fledged system with complete formal proofs for all reuse cases, optimized cache operations, and a complete negotiation framework.
We also plan to extend \textsc{ReBound} to richer forms of partial reuse and negotiation, and to extend it to support different query types.

\clearpage

\bibliographystyle{ACM-Reference-Format}
\bibliography{references}

\end{document}